\documentclass[pra,aps,superscriptaddress]{revtex4}

\usepackage{amsmath}
\usepackage{amsfonts}
\usepackage{amssymb}
\usepackage{amsthm}
\usepackage{bbm}
\usepackage{graphicx}
\usepackage{subfigure}

\newcommand{\idmat}{\mathbbm{1}}
\newcommand{\hs}{\mathcal{H}}
\newcommand{\trop}{\operatorname{Tr}}
\newcommand{\idop}{\operatorname{id}}
\newcommand{\beq}{\begin{equation}}
\newcommand{\eeq}{\end{equation}}
\newcommand{\ben}{\begin{eqnarray}}
\newcommand{\een}{\end{eqnarray}}

\newcommand{\ket}[1]{|#1 \rangle}
\newcommand{\bra}[1]{\langle #1 |}
\newcommand{\braket}[2]{\langle #1 |#2 \rangle}
\newcommand{\pro}[1]{\ket{#1}\bra{#1}}
\newcommand{\rmA}{{\rm A}}
\newcommand{\rmB}{{\rm B}}
\newcommand{\rmAB}{{\rm AB}}
\newcommand{\AtoB}{{\rmA\rightarrow\rmB}}
\newcommand{\BtoA}{{\rmA\leftarrow\rmB}}
\newcommand{\AntoB}{{\rmA\nrightarrow\rmB}}
\newcommand{\BntoA}{{\rmA\nleftarrow\rmB}}

\newtheorem{theorem}{Theorem}
\newtheorem{proposition}[theorem]{Proposition}
\newtheorem{definition}[theorem]{Definition}

\begin{document}

\title{On quantum non-signalling boxes}
\author{M. Piani}
\affiliation{Institute of Theoretical Physics and Astrophysics\\
University of Gda\'nsk, 80–952 Gda\'nsk, Poland}
\author{M. Horodecki}
\affiliation{Institute of Theoretical Physics and Astrophysics\\
University of Gda\'nsk, 80–952 Gda\'nsk, Poland}
\author{P. Horodecki}
\affiliation{Faculty of Applied Physics and Mathematics,\\
Technical University of Gda\'nsk, 80–952 Gda\'nsk, Poland}
\author{R. Horodecki}
\affiliation{Institute of Theoretical Physics and Astrophysics\\
University of Gda\'nsk, 80–952 Gda\'nsk, Poland}

\begin{abstract}
A classical non-signalling (or causal) box  is an operation on classical bipartite input with classical bipartite output such that no signal can be sent from a party to the other through the use of the box. The quantum counterpart of such boxes, i.e. completely positive trace-preserving maps on bipartite states, though studied in literature, have been investigated less intensively than classical boxes. We present here some results and remarks about such maps. In particular, we analyze: the relations among properties as causality, non-locality and entanglement; the connection between causal and entanglement breaking maps; the characterization of causal maps in terms of the classification of states with fixed reductions. We also provide new proofs of the fact that every non-product unitary transformation is not causal, as well as for the equivalence of the so-called semicausality
and semilocalizability properties.

\end{abstract}

\maketitle

\section{Introduction}
\label{sec:intro}
Non-relativistic quantum mechanics and special relativity coexist peacefully.
Namely, local measurements performed on spatially separated systems cannot be used to transmit messages, i.e. to signal.
To achieve signalling, communication - sending physical objects - is needed.  On the other hand, local
measurements can lead to nonlocal correlations, which though do not allow signalling, cannot be explained by 
any model that bases on local hidden variables. 

Suppose that Alice and Bob, who hold the two distant systems, communicate. Then stronger correlations 
than those exhibited by local measurements can be obtained. In general, this observation is rather obvious,
as communication is a nonlocal action. Yet, the issue become nontrivial, if one realizes that there are 
some operations on bipartite system, that need communication to be performed, but themselves 
cannot be used to communicate \cite{PR-machine,beckman}. Thus if Alice and Bob instead of being allowed to communicate, 
are allowed to apply such operations, for sure cannot use them to send signals.  For such operations, 
called in literature {\it non-signalling} or {\it causal boxes}, the 
following question arises: how strong can be the correlations exhibited by them? It turns out that 
even though they do not allow signalling, they can exhibit stronger violation of Bell's inequalities 
than local measurements performed on quantum states. This happens even in classical theory. 
An example is the PR-machine \cite{PR-machine}, which cannot signal itself, but needs (up to) one bit of communication
to be performed, and can violate CHSH inequalities in the maximal possible way admitted by the algebraic form 
of the inequalities (stronger than the maximal violation predicted for local measurements on quantum 
systems). Furthermore, the PR-machine can be used to reduce communication complexity of any one-bit valued function
down to one bit of communication \cite{vanDam-compl}. 
Thus the constraints given by non-signalling conditions  are not so stringent as it could seem, 
and they identify a highly nontrivial  set of operations both in classical and quantum information theory. 
It is therefore important to investigate the properties of non-signalling operations in order to 
understand interrelations between such notions as causality, correlations, nonlocality, communication,
communication complexity etc. In particular, a basic task is the description of the set of operations 
that cannot be used to communicate, but need communication to be performed. 

Interesting results have been obtained in \cite{PR-machine,beckman,semicausal,schumacher}. 
To be more specific, consider several notions introduced in 
\cite{beckman}. Maps that can be performed by local operations and bipartite ancilla in any needed state,
are called {\it localizable}. Maps that do not lead to communication are called {\it causal}
or {\it non-signalling}. Maps that can be performed by means of bipartite ancilla and at most communication 
in one direction are {\it semilocalizable}. Finally, maps that can be used to signal 
at most in one direction are called {\it semicausal}. In \cite{beckman} a characterization of the set of semicausal (hence also causal) 
maps was provided.
It was also shown that there are causal maps that are not localizable (which we have mentioned above: such maps need 
communication to be performed, but do not allow communication themselves). This result 
was actually already present in \cite{PR-machine}, 
where the authors showed that their box, the PR-machine, though causal, can violate 
Bell inequalities to a larger extent than quantum states do, which implies that it is not localizable in the language 
of \cite{beckman}.
In \cite{semicausal}
it was shown that semicausality and semilocalizability are equivalent. Other interesting questions 
concern the simulation of correlations obtained from maximally entangled states \cite{CerfGMP-simulating}.
Namely, it has been shown that a classical PR-machine (which is a resource weaker than one bit of communication)
can simulate results of local measurements performed on a singlet state.

In this paper we provide further results on non-signalling boxes, as well as new proofs 
of earlier results.  We  show that causal boxes are rare in the set of all maps. Then we 
investigate the connection between causal maps that are not localizable and the set of entanglement 
breaking maps. We link the classification of causal maps to the classification of states with same reductions. Subsequently, 
we study the trade-off between nonlocality and entangling power for a family of non-signalling maps. 
We also relate the issue of irreversibility for causal boxes  to a similar one in entanglement theory. 
We interpret the result of \cite{vanDam-compl}, that using non-signalling boxes  distant parties can compute any one-bit valued function of distributed data 
by use of one bit of communication, in terms of thermodynamical-like reversibility,
that was sought to present entanglement theory \cite{thermo-ent2002}. 

As far as new proofs are concerned, we provide a compact proof of 
the theorem of \cite{beckman} characterizing semicausal maps, of the fact that nonlocal unitaries 
are signalling, and that semicausal and semilocalizable maps are equivalent.

The paper is organized as follows. In Section \ref{sec:def} we recall the mathematical definition of causality and localizability, and provide some basic results. In Section \ref{sec:unitary} we focus on properties of unitary transformations, giving a new proof that non-factorized unitary transformations are non-causal, i.e. they allow communication.  Section \ref{sec:semilocal} is devoted to the equivalence of semicausality and semilocalizability. In Section \ref{sec:classes} we study some classes of causal maps; we define for causal maps their reduced maps, depicting a classification in classes of equivalence parallel to the ones for states with same reductions.In Section \ref{sec:classical} the relation between quantum causal maps and classical non-signalling boxes is investigated.  In Section \ref{sec:locality} we analyze the relationship between the standard quantum non-locality related to entanglement and the property of non-localizability of some causal maps. The communication irreversibility exhibited by boxes is discussed in analogy to bound entanglement in Section \ref{sec:bound}.

\section{Definitions and basic considerations}
\label{sec:def}
\subsection{Definitions}
In this section we provide basic definitions. Then we prove some properties of causal 
maps. In particular, we show that they are rare in the set of all physical maps, 
and provide a new short proof of characterization of causal maps of \cite{beckman}.

Completely positive trace-non increasing maps (superoperators) from density operators to density operators,
on Hilbert space $\hs$,
\beq
\label{eq:genmap}
\Lambda[\rho]=\sum_iK_i\rho K_i^\dagger,\qquad\sum_i K_i^\dagger K_i=\idmat,
\eeq
are the standard tool to describe the change (e.g. due to temporal evolution or to 
measurement) a quantum system undergoes. 
 We will now consider such maps on a bipartite system A+B with associated Hilbert 
space $\hs_{\rmAB}=\hs_{\rmA}\otimes\hs_{\rmB}$. We denote by $S_\rmA$, $S_\rmB$ and $S_\rmAB$ the 
sets of states of system A, B and A+B, respectively.

The causality properties of a map regard whether the map allows communication or not.
\begin{definition}
A map $\Lambda:S_\rmAB\rightarrow S_\rmAB$ is \emph{$\AntoB$ semicausal} if for every state $\rho_\rmAB$ and every map $\Gamma_\rmA:S_\rmA\rightarrow S_\rmA$ we have
\beq
\trop_\rmA\big(\Lambda[\rho_\rmAB]\big)=\trop_\rmA\big(\Lambda[(\Gamma_\rmA\otimes\idop)[\rho_\rmAB]]\big),
\eeq
that is no action of Alice before the global operation $\Lambda$ has any detectable consequence on Bob's side.

A similar definition holds for a map which is \emph{$\BntoA$ semicausal}. A map is  \emph{causal} if it is both $\AntoB$ and $\BntoA$ semicausal.

If a map is not $\AntoB$ semicausal, we say it is \emph{$\AtoB$ signalling}. If a map is not causal, we say it is \emph{signalling}.
\end{definition}

The localizability properties of a map regard whether the map needs communication (either classical or quantum) to be realized. 

\begin{definition}
A map $\Lambda:S_\rmAB\rightarrow S_\rmAB$ is \emph{$\AtoB$ semilocalizable} if the transformation it describes can be performed by one-way (either quantum or classical) communication from A to B, i.e. 
\beq
\Lambda[\rho_\rmAB]=\trop_R\big(G_{\rm BR} \circ F_{\rm AR}[\rho_\rmAB\otimes\omega_{\rm R}]\big),
\eeq
where $R$ is an ancilla system and $\circ$ denotes composition.

An operation is \emph{localizable} if it can be performed locally by sharing an ancilla system and no communication at all, i.e.
\beq
\Lambda[\rho_\rmAB]=\trop_{\rm RS}\big(F_{\rm AR}\otimes G_{\rm BS}[\rho_\rmAB\otimes\omega_{\rm RS}]\big)
\eeq
\end{definition}

It is immediate to check that $\AtoB$ semilocalizability implies $\BntoA$ semicausality, and therefore localizability implies causality. 
It is also evident~\cite{beckman} that semicausal (semilocalizable) maps form a convex set, therefore the same holds for causal maps, whose set is given by the intersection of the two convex sets of $\AntoB$ and $\BntoA$ semicausal maps. Also the set of localizable maps is convex~\cite{beckman}, but to show this is in principle less obvious, exactly because there are maps which, being causal, are both $\AtoB$ and $\BtoA$ semilocalizable, but are not localizable.  

\subsection{Causal maps are rare}
The following proposition asserts that semicausal maps are ``rare'' in the set of all CPT maps, because the smallest combination of $\AtoB$ signalling map with a $\AntoB$ semicausal map ruins the semicausality of the latter.
\begin{proposition}
\label{pro:zeromeasure}
The set of semicausal maps has an empty algebraic interior \footnote{See Appendix \ref{app:convex}.}.
\end{proposition}
\begin{proof}
Given a $\AntoB$ semicausal map $\Lambda$ and a $\AtoB$ signalling map $\Sigma$ the map $\Lambda(p)=p\Sigma+(1-p)\Lambda$ is $\AtoB$ signalling for any $0<p\leq1$. Indeed, consider a state $\rho_\rmAB$ and a map $\Gamma_\rmA$ such that 
\beq
\trop_\rmA\big(\Sigma[\rho_\rmAB]\big)\neq\trop_\rmA\big(\Sigma[(\Gamma_\rmA\otimes\idop)[\rho_\rmAB]]\big).
\eeq
Then, for $0<p\leq1$,
\beq
\begin{aligned}
\trop_\rmA\big(\Lambda(p)[(\Gamma_\rmA\otimes\idop)[\rho_\rmAB]]\big)
&=
p\trop_\rmA\big(\Sigma[(\Gamma_\rmA\otimes\idop)[\rho_\rmAB]]\big)+
(1-p)\trop_\rmA\big(\Lambda[(\Gamma_\rmA\otimes\idop)[\rho_\rmAB]]\big)\\
&=
p\trop_\rmA\big(\Sigma[(\Gamma_\rmA\otimes\idop)[\rho_\rmAB]]\big)+
(1-p)\trop_\rmA\big(\Lambda[\rho_\rmAB]\big)\\
&\neq
p\trop_\rmA\big(\Sigma[\rho_\rmAB]\big)+
(1-p)\trop_\rmA\big(\Lambda[\rho_\rmAB]\big)\\
&=\trop_\rmA\big(\Lambda(p)[\rho_\rmAB]\big),
\end{aligned}
\eeq
i.e. $\Lambda(p)$ is signalling.
\end{proof}

As a consequence, the set of semicausal maps has zero measure with respect to any reasonable measure on the set of maps.

One could wonder whether the previous theorem says that the set of non-signalling 
maps is a face of all CPT maps, i.e. whether the fact that a causal map $\Lambda_{\rm C}$ can be written as a non-trivial convex combination of two maps $\Lambda_1$ and $\Lambda_2$ forces also both the latter to be causal. It is not so. For example let us consider the totally depolarizing channel $D_\rmAB[X]=\trop[X]\idmat/(d_\rmAB)$, which is clearly causal. Fixing a basis $\{\ket{0},...\ket{d_\rmAB-1}\}$ in $\hs_{\rmAB}$, let us define the unitary operators
\beq
L\ket{j}=\ket{(j+1){\rm mod} (d_\rmAB)}\quad\textrm{and}\quad F\ket{j}=e^{2\pi ij/{d_\rmAB}}\ket{j}.
\eeq
Then $D_\rmAB[X]$ can be written as
\beq
D_\rmAB[X]=\frac{1}{(d_\rmAB)^2}\sum_{k=1}^{d_\rmAB}\sum_{l=1}^{d_\rmAB}L^k F^l X {F^l}^\dagger {L^k}^\dagger.
\eeq
Since the unitary operators $L^k F^l$ are in general entangling and thus signalling (see Section \ref{sec:unitary}), a causal superoperator as $D_\rmAB$ may be written as the convex combination of signalling maps.

\subsection{Characterization of semicausal maps}

We present here a compact proof of Theorem 8 of \cite{beckman}
characterizing the set of semicausal maps. 

\begin{theorem}
\label{th:semicausal}
A map $\Lambda$ is $\AntoB$ semicausal if and only if
\beq
\label{eq:condsemicausal}
\trop_\rmA\circ \Lambda\circ (D_\rmA\otimes\idop) = \trop_\rmA\circ \Lambda.
\eeq
\end{theorem}

\begin{proof}
(if) Obvious, since the totally depolarizing channel $D$ has the property $D\circ\Gamma=D$ for every trace-preserving map $\Gamma$.\\
(only if) By the definition of semicausality, we have that for every $\rho_\rmAB$ and $\Gamma_\rmA$:
\beq
\trop_\rmA\big(\Lambda[\rho_\rmAB]\big)=\trop_\rmA\big(\Lambda[(\Gamma_\rmA\otimes\idop)[\rho_\rmAB]]\big).
\eeq
Therefore we must have
\beq
\trop_\rmA\circ \Lambda=\trop_\rmA\circ \Lambda\circ (\Gamma_\rmA\otimes\idop),
\eeq
for every $\Gamma_\rmA$, in particular for $\Gamma_\rmA=D_\rmA$. This ends the proof.
\end{proof}

Condition \eqref{eq:condsemicausal} can be rewritten as
\beq
\label{eq:condsemicausal2}
(D_\rmA\otimes\idop)\circ \Lambda\circ (D_\rmA\otimes\idop) = (D_\rmA\otimes\idop)\circ \Lambda.
\eeq

Equality \eqref{eq:condsemicausal2} can be checked through the Choi-Jamio{\l}kowski isomorphism~\cite{choi,Jamiolkowski}: given a map $\Gamma_\rmAB$, we associate to it the unique operator
\beq
\rho_{\rm ABA'B'}=(\idop_{\rm A'B'}\otimes\Gamma_\rmAB)[P^+_{\rm AA'}\otimes P^+_{\rm BB'}],
\eeq
with $P^+_{\rm AA'}=\pro{\psi_{\rm AA'}^+}$, $\ket{\psi_{\rm AA'}^+}=1/\sqrt{d_\rmA}\sum_{i=1}^{d_\rmA} \ket{i}_\rmA\otimes\ket{i}_{\rm A'}$, and similarly for $P^+_{\rm BB'}$.

Remarkably, it follows that if a map is signalling, then a possible procedure for signalling is the following.
One takes the state violating the equality (\ref{eq:condsemicausal}) (it is easy to see 
that one can always chose it to be pure product one). Then Alice encodes ``0'' by doing nothing,
and she encodes ``1'' by depolarizing her system (replaces her system with one in the maximally mixed state). 
Then Alice and Bob apply the map, and by theorem, the two situations lead to different 
density matrix of Bob's system, which creates a channel of nonzero capacity from Alice to Bob. 

Condition \eqref{eq:condsemicausal2} can be used to check the semigroup property proved in~\cite{beckman}, i.e. that if $\Lambda_{1,2}$ are two $\AntoB$ semicausal maps, then also $\Lambda_1\circ\Lambda_2$ is $\AntoB$ semicausal. In fact,
\beq
\begin{split}
 (D_\rmA\otimes\idop)\circ (\Lambda_1\circ\Lambda_2)&=\big((D_\rmA\otimes\idop)\circ \Lambda_1\big)\circ\Lambda_2\\
 			&=\big((D_\rmA\otimes\idop)\circ \Lambda_1\circ(D_\rmA\otimes\idop)\big)\circ\Lambda_2\\
 			&=\big((D_\rmA\otimes\idop)\circ \Lambda_1\big)\circ\big((D_\rmA\otimes\idop)\circ\Lambda_2\big)\\
 			&=(D_\rmA\otimes\idop)\circ (\Lambda_1\circ\Lambda_2)\circ (D_\rmA\otimes\idop)
\end{split}
\eeq

\subsection{Reduced maps and states with fixed marginals}

\begin{figure}
\includegraphics[width=0.55\textwidth]{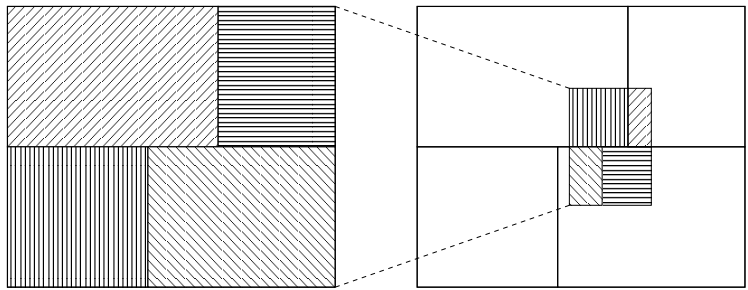}
\caption{Pictorial representation of causality: every whole convex equivalence class $[[\rho_\rmA\otimes\rho_\rmB]]$ (differently shaded regions) is mapped into a subset of an equivalence class.}
\label{fig:causalmap}
\end{figure}

Theorem \ref{th:semicausal} says immediately that if $\Lambda_\rmAB$ is $\AntoB$ semicausal, then the reduced density matrix $\rho'_\rmB=\trop(\rho'_\rmAB)=\trop(\Lambda[\rho'_\rmAB])$ depends only on $\rho_\rmB=\trop(\rho_\rmAB)$. This means that
an $\AntoB$ semicausal map $\Lambda_\rmAB$ maps two density matrices $\rho_\rmAB$ and $\sigma_\rmAB$ such that $\rho_\rmB=\sigma_\rmB$ into density matrices $\rho'_\rmAB$ and $\sigma'_\rmAB$ such that $\rho'_\rmB=\sigma'_\rmB$. Let us divide the set of states $S_\rmAB$ into the following convex classes of equivalence:
\begin{align}
[\rho_\rmA]_\rmA&\equiv\{\sigma_\rmAB|\sigma_\rmA=\rho_\rmA\},\\
[[\rho_\rmAB]]_\rmA&\equiv[\trop_\rmB(\rho_\rmAB)]_\rmA;
\end{align}
we define similarly $[\rho_\rmB]_\rmB$ and $[[\rho_\rmAB]]_\rmB$. We further define
\beq
[[\rho_\rmAB]]_\rmAB\equiv[[\rho_\rmAB]]_\rmA\cap[[\rho_\rmAB]]_\rmB,
\eeq
and, given a set of operators $S$, we denote by
\beq
\Lambda(S)=\{\Lambda(\sigma)|\sigma\in S\}
\eeq
the image of $S$ through a map $\Lambda$. The classes of equivalence $[[\rho_\rmA\otimes\rho_\rmB]]_\rmAB$ have been studied in \cite{partha,rudolph}; in particular their extremal point have been partly characterized. We notice that the property of causality of maps is strictly related to such a classification. Indeed, we have
\begin{proposition}
\label{pro:pro2}
A map $\Lambda_\rmAB$ is $\AntoB$ semicausal if and only if
\beq
\Lambda_\rmAB\big([[\rho_\rmAB]]_\rmB\big)\subseteq\big[\big[\Lambda_\rmAB(\rho_\rmAB)\big]\big]_\rmB,
\eeq
while it is causal if and only if (see Figure \ref{fig:causalmap})
\beq
\Lambda_\rmAB\big([[\rho_\rmAB]]_\rmAB\big)\subseteq\big[\big[\Lambda_\rmAB(\rho_\rmAB)\big]\big]_\rmAB.
\eeq
\end{proposition}
Given an $\BntoA$ semicausal map $\Lambda_\rmAB$ we can therefore consistently define the reduced map
\beq
\label{eq:redmap}
\Lambda_\rmA[\rho_\rmA]=\trop_\rmB(\Lambda_\rmAB[\rho_\rmAB])\equiv\trop_\rmA\left(\Lambda_\rmAB\left[\rho_\rmB\otimes\frac{\idmat_\rmB}{d_\rmB}\right]\right),
\eeq
with the choice $\rho_\rmB=\idmat_\rmB/d_\rmB$ arbitrary and irrelevant.
For semicausal maps (i.e. for maps for which the reduced map is well defined) we can then introduce classes parallel to the ones for states:
\begin{align}
[\Gamma_\rmA]_\rmA&\equiv\{\Omega_\rmAB|\Omega_\rmAB\textrm{ is $\BntoA$ semicausal},\,\Omega_\rmA=\Gamma_\rmA\},\\
[[\Lambda_\rmAB]]_\rmA&\equiv[\Lambda_\rmA]_\rmA,
\end{align}
with $\Lambda_\rmAB$ $\BntoA$ semicausal and $\Omega_\rmA$ and $\Lambda_\rmA$ reduced maps. For causal maps we can define
\beq
[[\Lambda_\rmAB]]_\rmAB\equiv[\Lambda_\rmA]_\rmA\cap[\Lambda_\rmB]_\rmB.
\eeq

Like the set $S_\rmAB$ is partitioned in convex subsets $[[\rho_\rmA\otimes\rho_\rmB]]_\rmAB$, $\rho_\rmA\in S_\rmA$, $\rho_\rmB\in S_\rmB$, the set of causal maps is partitioned in convex sets $[[\Lambda_\rmA\otimes\Lambda_\rmB]]_\rmAB$; without communication, neither Alice nor Bob can distinguish two states in the same $[[\rho_\rmA\otimes\rho_\rmB]]_\rmAB$ or the action of two maps in the same $[[\Lambda_\rmA\otimes\Lambda_\rmB]]_\rmAB$.

\section{Non-product unitary transformations are signalling: alternative proof}
\label{sec:unitary}

It has already been proved~\cite{beckman,bennett} that unitary maps $X\rightarrow U_\rmAB X U_\rmAB$ such that $U_\rmAB\neq U_\rmA\otimes U_\rmB$ are signalling. We provide here an alternative proof, which we consider to be simpler and more direct.

We notice that every unitary operator $U_\rmAB:\hs_{\rmAB}\rightarrow\hs_{\rmAB}$ such that for all $\ket{\psi_\rmA \otimes\psi_\rmB}\in\hs_\rmAB$ we have
\beq
U_\rmAB\ket{\psi_\rmA \otimes\psi_\rmB}=\ket{\psi'_\rmA \otimes\psi'_\rmB}
\eeq
has one of the following forms~\cite{busch}:
\begin{itemize}
\item[(a)] $U_\rmAB=U_\rmA\otimes U_\rmB$;
\item[(b)] $U_\rmAB\ket{\psi_\rmA \otimes\psi_\rmB}=V_{21}\ket{\psi_\rmB}\otimes W_{12}\ket{\psi_\rmA}$, where $V_{21}:\hs_\rmB\rightarrow\hs_\rmA$ and $W_{12}:\hs_\rmA\rightarrow\hs_\rmB$ are surjective isometries.
\end{itemize}
Case (b) can only occur if $\dim\hs_\rmA=\dim\hs_\rmB$.

It is evident that all unitaries corresponding to case (b) are signalling, while in case (a) signalling is not possible. Therefore we focus on unitary transformations with entangling power, i.e. there is a separable state $\ket{\psi_\rmA \otimes\psi_\rmB}\in\hs_\rmAB$ such that $U_\rmAB\ket{\psi_\rmA \otimes\psi_\rmB}$ is entangled.

Fix a state $\ket{\psi_\rmA}\in\hs_\rmA$. We have $\BntoA$ semicausality only if
\beq
U_\rmAB\ket{\psi_\rmA\otimes\psi_\rmB}=\sum_{i=1}^r\lambda_i\ket{i_\rmA\otimes i_\rmB},
\eeq
$\lambda_i>0$ for $i=1,...,r$, $\sum_{i=1}^r\lambda_i^2=1$, with the same Schmidt number $r$ and the same set $\{(\lambda_i,\ket{i_\rmA})\}$ for any arbitrary state $\ket{\psi_\rmB}\in\hs_\rmB$, so that for any possible choice of $\ket{\psi_\rmB}$ on side B, the reduction $\rho_A$ after the action of the unitary transformation is the same. Indeed, according to~\cite{josza} if we consider two purifications $\ket{\xi_1}_\rmAB$ and $\ket{\xi_2}_\rmAB$ of the same state $\rho_A$, they are linked by a local unitary transformation acting on B; in our case, we must therefore have
\beq
U_\rmAB\ket{\phi\otimes\psi'}=(\idmat\otimes V_\rmB)U_\rmAB\ket{\phi\otimes\psi}
\eeq
if we want to exclude $\BtoA$ signalling. A similar request holds to forbid $\AtoB$ signalling.

The previous requirement is not satisfied by any entanglement creating unitary operation $U_\rmAB$.
Let us fix a state $\ket{\phi}$ in $\hs_\rmA$ and an orthonormal basis $\{\ket{\psi_a}\}$ in $\hs_\rmB$.
As just explained, to satisfy the non-signalling condition it must be
\beq
U_\rmAB\ket{\phi\otimes\psi_a}=\sum_{i=1}^r\lambda_i\ket{i_\rmA\otimes i_\rmB^a},
\eeq
with $r$ the Schmidt number and the dependence of the Schmidt decomposition on $\ket{\psi_a}$ appearing only on side B.
The reduced density matrix obtained tracing over the B degrees of freedom after the action of $U_\rmAB$ is
\beq
\rho_A=\sum_{i=1}^r\lambda_i\pro{i_\rmA}.
\eeq
Let us now consider the action of $U_\rmAB$ on the separable state $\ket{\phi\otimes(\alpha\psi_a+\beta\psi_b)}$, $a\neq b$, with 
\beq
\label{eq:normalization}
|\alpha|^2+|\beta|^2=1,
\eeq
i.e.
\beq
U_\rmAB\ket{\phi\otimes(\alpha\psi_a+\beta\psi_b)}=
\sum_{i=1}^r\lambda_i
\ket{i_\rmA}
\otimes
(\alpha\ket{i_\rmB^a}+\beta\ket{i_\rmB^b}).
\eeq
In this case we have
\beq
\begin{split}
\rho'_\rmA&=
\sum_{i=1}^r\lambda_i^2\pro{i_\rmA}\\
&+
\alpha\beta^*
\sum_{i,j=1}^r\lambda_i\lambda_j
\ket{i_\rmA}\bra{j_\rmA}
\braket{j_\rmB^b}{i_\rmB^a}
+
\alpha^*\beta
\sum_{i,j=1}^r\lambda_i\lambda_j
\ket{i_\rmA}\bra{j_\rmA}
\braket{j_\rmB^a}{i_\rmB^b}.
\end{split}
\eeq
For the transformation to be non-signalling $\rho'_\rmA=\rho_\rmA$; because of the freedom in the choice of $\alpha,\,\beta$ it must be
\begin{align}
\sum_{i,j=1}^r\lambda_i\lambda_j
\ket{i_\rmA}\bra{j_\rmA}
\braket{j_\rmB^b}{i_\rmB^a}&=0\\
\sum_{i,j=1}^r\lambda_i\lambda_j
\ket{i_\rmA}\bra{j_\rmA}
\braket{j_\rmB^a}{i_\rmB^b}&=0,
\end{align}
i.e. 
\beq
\braket{j_\rmB^b}{i_\rmB^a}=0
\eeq
for $a\neq b$ and $i,j=1,...,r$. Since each $\{\ket{i_\rmB^a}\}_{i=1}^r$ is already an orthonormal set, we have
\beq
\braket{j_\rmB^b}{i_\rmB^a}=\delta_{ij}\delta_{a,b}
\eeq
with $i,j=1,...,r$ and $a,b=1,...,d_\rmB$. Therefore there should be $rd_\rmB$ orthonormal states in $\hs_\rmB$, which is possible only if $r=1$, i.e. if the unitary transformation does not create entanglement.
This ends the proof.

It is worth noting that as regards the partitioning of causal maps in convex sets $[[\Lambda]]_\rmAB$, we have $[[U_\rmA\otimes U_\rmB]]=\{U_\rmA\otimes U_\rmB\}$, since it is readily seen that
\begin{align}
\trop_\rmB(\Lambda[\pro{\psi}\otimes\pro{\phi}])&=U_\rmA\pro{\psi}U_\rmA^\dagger\\
\trop_\rmA(\Lambda[\pro{\psi}\otimes\pro{\phi}])&=U_\rmB\pro{\psi}U_\rmB^\dagger
\end{align}
implies $\Lambda[X]=U_\rmA\otimes U_\rmB\,X\,U_\rmA^\dagger\otimes U_\rmB^\dagger$.

\section{Semicausality is equivalent to semilocalizability: new proof}
\label{sec:semilocal}

That semicausality implies semilocalizability (the converse implication being quite trivial) was proved in~\cite{semicausal}, using the uniqueness of the Stinespring representation for a completely positive map. Here we provide a slightly different reasoning leading to the same result, using the Choi-Jamio{\l}kowski isomorphism and Theorem \ref{th:semicausal}.

Given a $\AntoB$ semicausal map $\Lambda_\rmAB$we consider its isomorphic state
\beq
\rho_{\rm AA'BB'}=(\idop_{\rm A'B'}\otimes\Lambda_\rmAB)[P^+_{\rm AA'}\otimes P^+_{\rm BB'}].
\eeq
According to condition \eqref{eq:condsemicausal}, $\rho_{\rm AA'BB'}$ has reduction
\beq
\rho_{\rm A'BB'}=\trop_\rmA\big(\rho_{\rm AA'BB'}\big)=\frac{\idmat}{d_{{\rm A}'}}\otimes\rho_{\rm BB'},
\eeq
with
\beq
\rho_{\rm BB'}=\Lambda_\rmB\otimes\idop_{\rm B'}[P^+_{\rm BB'}]
\eeq
where $\Lambda_\rmB$ is the reduced map of $\Lambda_\rmAB$ as defined in \eqref{eq:redmap}. Since $\Lambda_\rmB$ is a CPT map,
\beq
\rho_{\rm BB'}=\trop_{\rm E}(U_{\rm BE}\otimes\idmat_{\rm B'}P^+_{\rm BB'}\otimes\pro{0_{\rm E}}U_{\rm BE}^\dagger\otimes\idmat_{\rm B'}),
\eeq
with E an auxiliary system of dimension at most $d_\rmB^2$ and $\ket{0_{\rm E}}$ a reference state of its.
Therefore $\rho_{\rm A'BB'}$ admits a purification of the form
\beq
\ket{\phi_{\rm AA'BB'E}}
=
\ket{\psi_{\rm AA'}^+}
\otimes
\big(
	(U_{\rm BE}\otimes\idmat_{\rm B'})
	(\ket{\psi_{\rm BB'}^+}\ket{0_{\rm E}}\big).
\eeq

Every purification $\ket{\psi_{\rm AA'BB'C}}$ of $\rho_{\rm AA'BB'}$ is also a purification of $\rho_{\rm A'BB'}$, thus it must be
\beq
\begin{aligned}
\ket{\psi_{\rm AA'BB'C}}
&=
(\idmat_{\rm A'BB'}\otimes U_{\rm AC})
\ket{\phi_{\rm AA'BB'C}}\\
&=
\bigg(
	(\idmat_{\rm A'BB'}\otimes U_{\rm AC})
	\circ
	(U_{\rm BC}\otimes\idmat_{\rm B'})
\bigg)
\Big(
	\ket{\psi_{\rm AA'}^+}\otimes\ket{\psi_{\rm BB'}^+}\otimes\ket{0}_{\rm C}
\Big)
\end{aligned}
\eeq
with E embedded in C, so that $\ket{\phi_{\rm AA'BB'E}}\mapsto\ket{\phi_{\rm AA'BB'C}}$, $U_{\rm BC}\mapsto U_{\rm BE}$ and $\ket{0_{\rm E}}\mapsto\ket{0_{\rm C}}$ , and $U_{\rm AC}$ a unitary operation on A and C only.

So we have found
\beq
\label{eq:semiequi}
(\idop_{\rm A'B'}\otimes\Lambda_\rmAB)[P^+_{\rm AA'}\otimes P^+_{\rm BB'}]
=
\trop_{\rm C}
\Big[
	(\idmat_{\rm A'BB'}
		\otimes (U_{\rm AC}\circ U_{\rm BC}))\big(P_{\rm AA'}^+\otimes P_{\rm BB'}^+\otimes\pro{0_{\rm C}}\big)(\idmat_{\rm A'BB'}
		\otimes (U_{\rm AC}\circ U_{\rm BC}))^\dagger\Big].
\eeq
Because of the Choi-Jamio{\l}kowski isomorphism, the right-hand side of \eqref{eq:semiequi}  defines a unique map $S_\rmAB\rightarrow S_\rmAB$, which is $\BtoA$ semilocalizable and coincides with the starting map $\Lambda_\rmAB$.

\section{Classes of causal maps}
\label{sec:classes}

In this section we discuss the relation between causal maps and entanglement breaking maps, 
as well as some positive but not completely positive maps. We provide also a general scheme to ``construct'' causal maps.

\subsection{Causal maps and entanglement-breaking maps}
Note that all known examples of non-trivial causal maps, in particular the ones that are not localizable, belong to the class of entanglement breaking trace-preserving (EBT) maps~\cite{EBT}:
\beq
\label{eq:nlm}
\Lambda[\rho_\rmAB]=\sum_i \trop(F_i\rho_\rmAB)\sigma^i_\rmAB,
\eeq
with $\{F_i\}$ a POVM, i.e. $F_i\geq0$, $\sum_iF_i=\idmat$.

We can apply Theorem \ref{th:semicausal} to this class of maps. We find that an EBT map $\Lambda$ is $\AntoB$ semicausal if and only if
\beq
\rho'_\rmB=\trop_\rmA[\Lambda[\rho_\rmAB]]=\sum_i \trop_B(\trop_\rmA[F_i/d_\rmA]\rho_\rmB)\sigma^i_\rmB.
\eeq
Such condition is clearly satisfied if for example
\beq
\label{eq:reductioncond}
\sigma_\rmB^i=\sigma_\rmB\;\forall i.
\eeq
or if $F_i=\idmat\otimes N_i$ with $\{N_i\}$ a POVM on the B subsystem only.

Notice that even if a map $\Lambda$ as in \eqref{eq:nlm} is an EBT map, i.e.
\beq
(\Lambda\otimes\idop_{\rm C})[\rho_{\rm ABC}]
\eeq
is always separable with respect to the cut AB$|$C, for any state $\rho_{\rm ABC}$, 
it may create entanglement between A and B, since the states $\sigma^i_\rmAB$ may be entangled.

\begin{figure}
\includegraphics[width=0.5\textwidth]{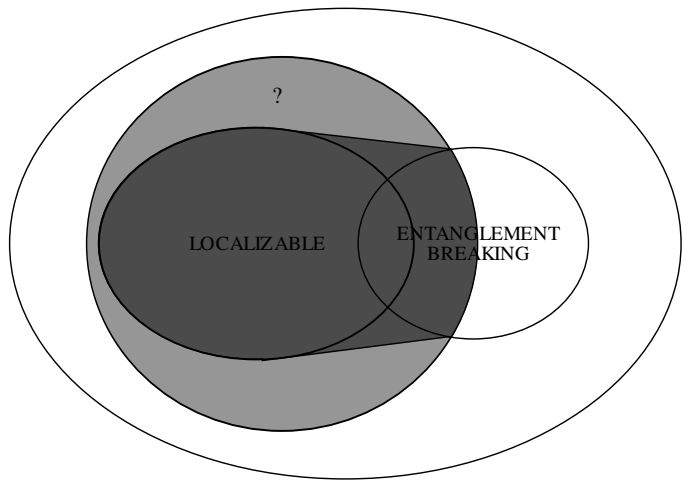}
\caption{Relations between the different sets of maps: entanglement breaking trace-preserving (EBT), localizable, causal (light grey), convex hull of localizable and causal EBT maps (dark grey). No example of a map which is not in the latter set but it is causal is known at present. Notice that because of Proposition \ref{pro:zeromeasure} any non-trivial convex combination of a localizable map and a non-causal EBT map is non-causal.}
\label{fig:sets}
\end{figure}

An interesting question is the following: is the set of all causal maps contained 
in the convex hull of localizable and causal EBT maps? If so, it would mean that non-trivial causal maps (e.g. non-localizable) have some ``classical'' feature: they destroy (completely, if EBT) quantum information contained in the initial state. The current knowledge about the set of causal maps is 
presented on Figure \ref{fig:sets}.

\subsection{Non-signalling maps from positive maps}

It is possible to find examples of non-signalling maps starting from a non-signalling map $\Lambda_{\rm P}$ which is positive but not completely positive and considering the convex combination $p\Lambda_{\rm P}+(1-p)D_\rmAB$ with $p$ such that the resulting map is completely positive~\cite{pawelekert}. Here are some examples:
\begin{itemize}
\item $\Lambda_{\rm P}=T$, the transposition with respect to a fixed basis: $0\leq p\leq\frac{1}{d_\rmAB+1}$;
\item $\Lambda_{\rm P}[X]=-X$, the total reflection: $0\leq p\leq\frac{1}{d_\rmAB^2+1}$;
\item in the case of two qubits, with $\rho_\rmAB=\sum_{\alpha,\beta=0}^3\rho_{\alpha,\beta}\sigma_\alpha\otimes\sigma_\beta$, $\Lambda_{\rm P}[\rho_\rmAB]=\sum_{\alpha,\beta=0}^3\xi_{\alpha\beta}\rho_{\alpha,\beta}\sigma_\alpha\otimes\sigma_\beta$, with $\xi_{\alpha\beta}=-1$ if $\alpha\neq0$ and $\beta\neq0$ and $\xi_{\alpha\beta}=1$ otherwise: $0\leq p\leq\frac{1}{3}$.
\end{itemize}

Unfortunately, such examples do not provide any more insight as regards the structure of causal maps, since they are all localizable; in particular they can been realized by shared randomness. In fact, any positive map $\Lambda$ admits a diagonal representation
\beq
\label{eq:diagexpress}
\Lambda[X]=\sum_i \lambda_i F_i X F_i^\dagger
\eeq
with $\{F_i\}$, $\trop(F_i^\dagger F_j)=\delta_{ij}$ an orthonormal operator basis in $B(\hs)$, the set of bounded operators on $\hs$, and $\lambda_{i}\in \mathbb{R}$. Note that 
given any orthonormal operator basis $\{F_i\}$ we have
\beq
D[X]=\sum_i F_i X F_i^\dagger=\trop(X)\idmat;
\eeq
moreover, given two orthonormal basis $\{A_i\}$ and $\{B_j\}$ for $B(\hs_\rmA)$ and $B(\hs_\rmB)$, respectively, $\{A_i\otimes B_j\}$ is a basis for $B(\hs_\rmAB)$. All the considered positive maps have can be expressed in a diagonal form
\beq
\label{eq:diagexpressAB}
\Lambda[X]=\sum_{i,j} \lambda_{ij}A_i\otimes B_j X A_i^\dagger\otimes B_j^\dagger,
\eeq
with $\lambda_{ij}\in \mathbb{R}$; for example, the global transposition is given by the composition of local transpositions, $T_\rmAB=T_\rmA\otimes T_\rmB$, with each $T_{\rm x}$, $x=A,B$, having the expression \eqref{eq:diagexpress}. Therefore every completely positive convex combination $p\Lambda_{\rm P}+(1-p)D_\rmAB$ has an expression \eqref{eq:diagexpressAB} with $\lambda_{ij}\geq0$.

\subsection{A general scheme to construct causal maps}

Given a $d$-dimensional Hilbert space $\hs_d$, it is possible to consider as a (non-orthonormal) basis for $B(\hs_d)$ a set of state, in particular of projectors on pure states $\{P_1,\ldots,P_{d^2}\}$, with $P_i=\pro{\psi_i}$. For example, given a basis $\{\ket{1},\ldots,\ket{d}\}$ in $\hs_d$, we can consider the operator basis given by the $d^2$ projectors on
\ben
&\ket{k} & k=1,\ldots,d\\ 
&\dfrac{\ket{k}+\ket{l}}{\sqrt{2}},\dfrac{\ket{k}+i\ket{l}}{\sqrt{2}} & k,l=1,\ldots,d;\,k>l.
\een
Since this is a basis, any trace-preserving linear map $\Lambda$ may be thought as completely defined by the assignment
\beq
\label{eq:mapassign}
\Lambda[P_i]=\rho_i,\quad i=1,\ldots,d^2.
\eeq
In general, with arbitrary choice of $\rho_i$, the resulting map is not only non-completely positive, but not even positive~\footnote{For example, in the case of qubit map, let us consider the operator basis given by projectors onto states $\ket{\psi_1}=\ket{0}$, $\ket{\psi_1}=\ket{1}$, $\ket{\psi_2}=(\ket{0}+\ket{1})/\sqrt{2}$, $\ket{\psi_2}=(\ket{0}+i\ket{1})/\sqrt{2}$; then the map $\Lambda$ such that $\Lambda[P_i]=P_{\sigma(i)}$, with $\sigma$ the permutation $(1\,3)$, is not positive because $\Lambda[P_1+P_2-P_3]=P_3+P_2-P_1\ngeq 0$.}. In any case, given an arbitrary assignment \eqref{eq:mapassign}, i.e. an arbitrary trace-preserving linear map, we can always construct a convex combination $p\Lambda+(1-p)D$ with $p$ such that the resulting map is completely positive.

In our case we can therefore consider an operator basis in $B(\hs_\rmAB)$ given by some separable product states $P^\rmA_1\otimes P^\rmB_1,\ldots,P^\rmA_{d_\rmA^2}\otimes P^\rmB_{d_\rmB^2}$. A trace-preserving linear map $\Lambda$ is defined by the assignment $\Lambda[P^\rmA_i\otimes P^\rmB_j]=\rho_\rmAB^{ij}$, $i=1,\ldots,d_\rmA^2$, $j=1,\ldots,d_\rmB^2$, with $\rho_\rmAB^{ij}\in S_\rmAB$. A sufficient and necessary requirement for the resulting map to be causal is given by Proposition \ref{pro:pro2} and in this case may be stated as
\beq
\trop_\rmB(\rho_\rmAB^{ij})=\rho_\rmA^i,\quad\trop_\rmA(\rho_\rmAB^{ij})=\rho_\rmB^j,
\eeq
i.e. $\rho_\rmAB^{ij}\in[[\rho_\rmA^i\otimes\rho_\rmB^j]]_\rmAB$ (see Table \ref{tab:causalmap}): the impossibility of sending signals by choosing different elements in $\{P^\rmA_i\otimes P^\rmB_j\}$ is necessary and sufficient for causality. 
\begin{table}
	\centering
		\begin{tabular}{c|ccccc}
						&	$\rho_\rmB^1$ 		& $\rho_\rmB^2$ 		& \ldots 	& $\rho_\rmB^{d_\rmB^2}$\\
\hline
$\rho_\rmA^1$	& $\rho_\rmAB^{11}$	& $\rho_\rmAB^{12}$	& \ldots	& $\rho_\rmAB^{1 d_\rmB^2}$\\
$\rho_\rmA^2$	& $\rho_\rmAB^{21}$	& $\rho_\rmAB^{22}$	& \ldots	& $\rho_\rmAB^{2 d_\rmB^2}$\\
\vdots			&			\vdots				&				\vdots				&	$\ddots$	&				\vdots				\\
$\rho_\rmA^{d_\rmA^2}$	& $\rho_\rmAB^{d_\rmA^21}$	& $\rho_\rmAB^{d_\rmA^22}$	& \ldots	& $\rho_\rmAB^{d_\rmA^2 d_\rmB^2}$
		\end{tabular}
	\caption{A non-signalling map may be defined by assigning $d_\rmA^2 d_\rmB^2$ states $\rho_\rmAB^{ij}=\Lambda[P^\rmA_i\otimes P^\rmB_j]$ such that $\trop_\rmB(\rho_\rmAB^{ij})=\rho_\rmA^i$, $\trop_\rmA(\rho_\rmAB^{ij})=\rho_\rmB^j$.}
	\label{tab:causalmap}
\end{table}
Note that fixing the reduced states $\{\rho_\rmA^i\}$ and $\{\rho_\rmB^j\}$ we define a unique map $\Lambda_\rmA\otimes\Lambda_\rmB$; we obtain the set $[[\Lambda_\rmA\otimes\Lambda_\rmB]]_\rmAB$varying each state $\rho_\rmAB^{ij}$ in $[[\rho_\rmA^i \otimes\rho_\rmB^j]]_\rmAB$ (and requiring the complete positivity of the resulting map).

\section{Relation with classical non-signalling boxes}
\label{sec:classical}

Classical non-signalling boxes take the input from two parties and furnish an output such that the parties can not communicate, i.e. Alice can not understand from her output what was Bob's input. Still, when they meet afterwards they may find that the recorded outputs are more correlated than in the case in which the output on one side was completely independent from the input on the other side. In this case we speak of non-signalling \emph{non-local} classical boxes, since the implementation of such boxes requires communication. 

Classical boxes are described by conditional probabilities $p_{ab|xy}$ of obtaining outputs $a$ and $b$ with inputs $x$ and $y$, on Alice's and Bob's side respectively~\cite{BLMPPR}. Such conditional probabilities form a convex set. In the case of two possible distinct inputs, $x,\,y\in\{0,1\}$, and $d_\rmA,\, d_\rmB$ possible distinct outputs $a\in\{0,1,...,d_\rmA-1\}$, $b\in\{0,1,...,d_\rmB-1\}$ on Alice's and Bob's side respectively, we have a characterization of the (non-local) extreme points of such a convex set (apart local reversible transformation, e.g. relabellings):
\beq
\label{eq:nlcb}
p_{ab|xy}=\begin{cases}
1/k & (b-a)\,{\rm mod}\,k=x\cdot y\,\land\,a,b\in\{0,...,k-1\}\\
0 & \textrm{otherwise}
\end{cases}
\eeq
for $k=\{2,...,\min(d_\rmA,d_\rmB)\}$.
This means that for fixed $k$, on each side the output is an apparently random number among $\{0,...,k-1\}$ even if the outputs are strictly correlated.

It is not difficult to check that a corresponding quantum non-local box may be given by the map:
\beq
\label{eq:qnlb}
\Lambda_k[\pro{\psi_{\rm in}}]
=\trop_{\rm in}\big(U(\pro{\psi_{\rm in}}\otimes\pro{\psi^k_{\rm a,b}})U^\dagger\big),
\eeq
with
\beq
\ket{\psi^k_{\rm a,b}}=\frac{1}{\sqrt{k}}\sum_{i=0}^{k-1}\ket{i_a\otimes i_b}
\eeq
and
\beq
U\ket{a\otimes b\otimes x\otimes y}=\ket{a\otimes (b\oplus_k x\cdot y) \otimes x\otimes y},
\eeq
where $m\oplus_k n=(m+n)\,{\rm mod}\,k$.
Indeed, ``feeding'' such map with a separable state $\ket{\psi_{\rm in}}=\ket{x\otimes y}$, $x,\,y\in\{0,1\}$, and measuring in the factorized basis $\ket{a\otimes b}$, $a\in\{0,1,...,d_\rmA-1\}$, $b\in\{0,1,...,d_\rmB-1\}$, we obtain exactly the conditional probabilities \eqref{eq:nlcb}. In general, with input $\rho_{\rm in}$, the action of the quantum box corresponds to
\beq
\label{eq:nl_qc_b}
\Lambda_k[\rho_{\rm in}]=(1-p)\pro{\psi^k_{\rm a,b}}+p(\idmat_\rmA\otimes\Sigma)\pro{\psi^k_{\rm a,b}}(\idmat_\rmA\otimes\Sigma^\dagger)
\eeq
with $p=\bra{11}\rho_{\rm in}\ket{11}$ and $\Sigma$ a permutation of order $k$ realizing the transformation $\Sigma\ket{b}=\ket{b\oplus_k 1}$.
Such map is therefore of the form \eqref{eq:nlm} (with the slight difference that input and output dimensions are different) and condition \eqref{eq:reductioncond} is valid both for A and B, so that the map is causal.

Another possible choice for a quantum version of the non-local box is given by
\beq
\Lambda'_k[\rho_{\rm in}]=(1-p)\pro{\psi^k_{\rm a,b}}+p(\idmat_\rmA\otimes\Sigma)\rho_{a,b}^k(\idmat_\rmA\otimes\Sigma^\dagger)
\eeq
with $\rho_{a,b}^k=\frac{1}{k}\sum_{i=0}^{k-1}\pro{i_a\otimes i_b}$. The two possible quantum versions differ in \emph{coherence}: in the second case local measurements are performed, and then the classical box is operated.
Note that the map $\Lambda_{\rm NL}$ can be realized either with classical communication having at disposal a shared maximally entangled state or with quantum communication. The map $\Lambda_{\rm NL}$ requires only classical communication, as the classical version.

In the next section we will focus on the case $k=d_\rmA=d_\rmB=2$, that is on the map
\beq
\label{eq:nl_qc_b_2}
\Lambda_{\rm NL}[\rho_{\rm in}]=(1-p)\pro{\psi_0}+p\pro{\psi_1}
\eeq
with, as before, $p=\bra{11}\rho_{\rm in}\ket{11}$ and
\beq
\ket{\psi_0}=\frac{\ket{00}+\ket{11}}{\sqrt{2}},\quad
\ket{\psi_1}=(\idmat\otimes\sigma_1)\ket{\psi_0}=\frac{\ket{01}+\ket{10}}{\sqrt{2}},
\eeq
where $\sigma_1=\begin{pmatrix}0&1\\1&0\end{pmatrix}$.
Its ``incoherent'' version was considered in \cite{beckman}:
\beq
\label{eq:nl_qc_b_2_prime}
\Lambda'_{\rm NL}[\rho_{\rm in}]=(1-p)\frac{\ket{00}\bra{00}+\ket{11}\bra{11}}{2}+p\frac{\ket{01}\bra{01}+\ket{10}\bra{10}}{2}.
\eeq
It was there proved that with such a transformation at disposal, it is possible to violate Bell inequalities up to their algebraic maximum, so that the map can not be localizable (if the map is localizable Cirel'son's bound holds~\cite{cirelson}). In the next section we analyze the interplay between such a non-localizability feature and the non-locality related to entanglement.

Notice that considering boxes as \emph{primitives}, i.e. as if they existed in nature so that we do not care about their realization, the coherent non-local quantum box $\Lambda_{\rm NL}$ can be used to \emph{produce} a singlet. Hence it can reproduce {\it all} phenomena exhibited by the latter state. This is in contrast with the classical PR-machine (or the incoherent quantum
box $\Lambda'_{\rm NL}$), which can reproduce results of local measurement performed on the singlet,
but cannot for example reproduce teleportation. Indeed, the classical box can be implemented by use of classical communication, which cannot convey quantum information as teleportation does.

\section{Entanglement, entangling power, non-locality and communication complexity}
\label{sec:locality}

We will consider a family of coherent maps, which exhibit a trade-off between the degree of nonlocality
measured by violation of the CHSH inequality and the average entanglement  created out of product states 
\cite{Zanardi-entpower}. We also compare the degree of maximal violation obtainable with coherent and incoherent versions of non-local boxes.

Let us consider the standard CHSH inequality, involving the quantity
\beq
I(a,a';b,b')=E(a,b)+E(a,b')+E(a',b)-E(a',b'),
\eeq
where the label $a,a'$ distinguish two possible setups for an experiment on A side, with each run of the experiment conveying one result $r_{a(a')}=\pm1$; $b,b'$ label similarly two possible setups on B side, and $E(c,d)$ stays for the expectation value of the product $r_cs_d$ of the result $r_c$ of experiment $c$ on side A and $s_d$ of experiment $d$ on side B.
Local hidden variables models obey
\beq
\label{eq:CHSH_piani}
I(a,a';b,b')\leq2,
\eeq
while when the expectation values are calculated according to quantum mechanics with respect to a fixed state $|\psi\rangle$, i.e.
$E(a,b)=\langle\psi | A_a\otimes B_b |\psi\rangle$, with $A_a$ and $B_b$ dichotomic observables for A and B respectively, (\ref{eq:CHSH_piani}) can be violated, that is one can have $I(a,a';b,b')>2$.
Anyway, there is a limit for the violation of (\ref{eq:CHSH_piani}) in this framework, the so-called Cirel'son bound~\cite{cirelson}, $I(a,a';b,b')\leq2\sqrt{2}$.

We consider two different experimental setups on side A which comprise the preparation of one between two possible orthogonal initial state $\ket{0}_\rmA,\ket{1}_\rmA$; each state corresponds to an observable $\vec{a}_{x}\cdot\vec{\sigma}$, $x=0,1$. Similarly for B, with observables $\vec{b}_{y}\cdot\vec{\sigma}$, $y=0,1$. The experimental procedure is the following:
\begin{enumerate}
\label{experiment}
\item Alice prepares a state $\ket{x}_\rmA$, with either $x=0$ or $x=1$; similarly for Bob, with $\ket{y}_\rmB$, $y=0,1$;
\item the state $\ket{\psi_{\rm in}}=\ket{x}_\rmA\otimes\ket{y}_\rmB$ is fed to the quantum non-local box \eqref{eq:nl_qc_b_2};
\item Alice and Bob measure observables $\vec{a}_{x}\cdot\vec{\sigma}$ and $\vec{b}_{y}\cdot\vec{\sigma}$ with respect to the output $\Lambda_k[\pro{\psi_{\rm in}}]$.
\end{enumerate}
It is easy to check that
\begin{align}
\bra{\psi_0}
\vec{a}_{x}\cdot\vec{\sigma}\otimes\vec{b}_{y}\cdot\vec{\sigma}\ket{\psi_0}&=\sum_i({a}_{x})_i({b}_{y})_i(-1)^{\delta_{i,2}}\\
\bra{\psi_1}
\vec{a}_{x}\cdot\vec{\sigma}\otimes\vec{b}_{y}\cdot\vec{\sigma}\ket{\psi_1}&=-\sum_i({a}_{x})_i({b}_{y})_i(-1)^{\delta_{i,1}+\delta_{i,2}}.
\end{align}
Choosing
\beq
\vec{a}_0=\vec{a}_1=\vec{b}_0=\vec{b}_1=(1,0,0)
\eeq
we obtain for the experimental procedure depicted above $I(\vec{a}_0,\vec{a}_1;\vec{b}_0,\vec{b}_1)=4$.

Contrary to its dephased version, the map \eqref{eq:nl_qc_b_2} has an entangling power, i.e. it can map separable states into entangled states. Therefore, it is not only non-localizable, but also gives raise to ``standard'' quantum non-locality. We will study the interplay between these two features considering the one-parameter map
\beq
\Lambda_\alpha[\rho_{\rm in}]=(1-\alpha p)\pro{\psi_0}+\alpha p\pro{\psi_1};
\eeq
for $\alpha=1$ we recover the map \eqref{eq:nl_qc_b_2}, while for $\alpha=0$ we get an operation that replaces the input state with $\pro{\psi_0}$, i.e. a constant map with maximally entangled output. Such an operation can be realized by Alice and Bob through the use, apart of the resources needed for $\Lambda_{\rm NL}$, of local shared randomness. They check their shared random number $\lambda\in [0,1]$: if $0\leq\lambda\leq\alpha$ they proceed with the protocol realizing $\Lambda_{\rm NL}$, otherwise their output is $\pro{\psi_0}$. The communication complexity of  implementing $\Lambda_\alpha$ is therefore (at most) $\alpha$ times the communication complexity of realizing $\Lambda_{\rm NL}$.

\subsection{Entanglement and communication complexity for the maximal violation of the CHSH inequality} 

\begin{figure}
\centering
\subfigure[$0\leq\alpha\leq \frac{2}{3},\,\varphi=\arcsin\sqrt{\frac{2-3\alpha}{4(1-\alpha)}}$]{\label{fig:optob}\includegraphics[width=0.35\textwidth]{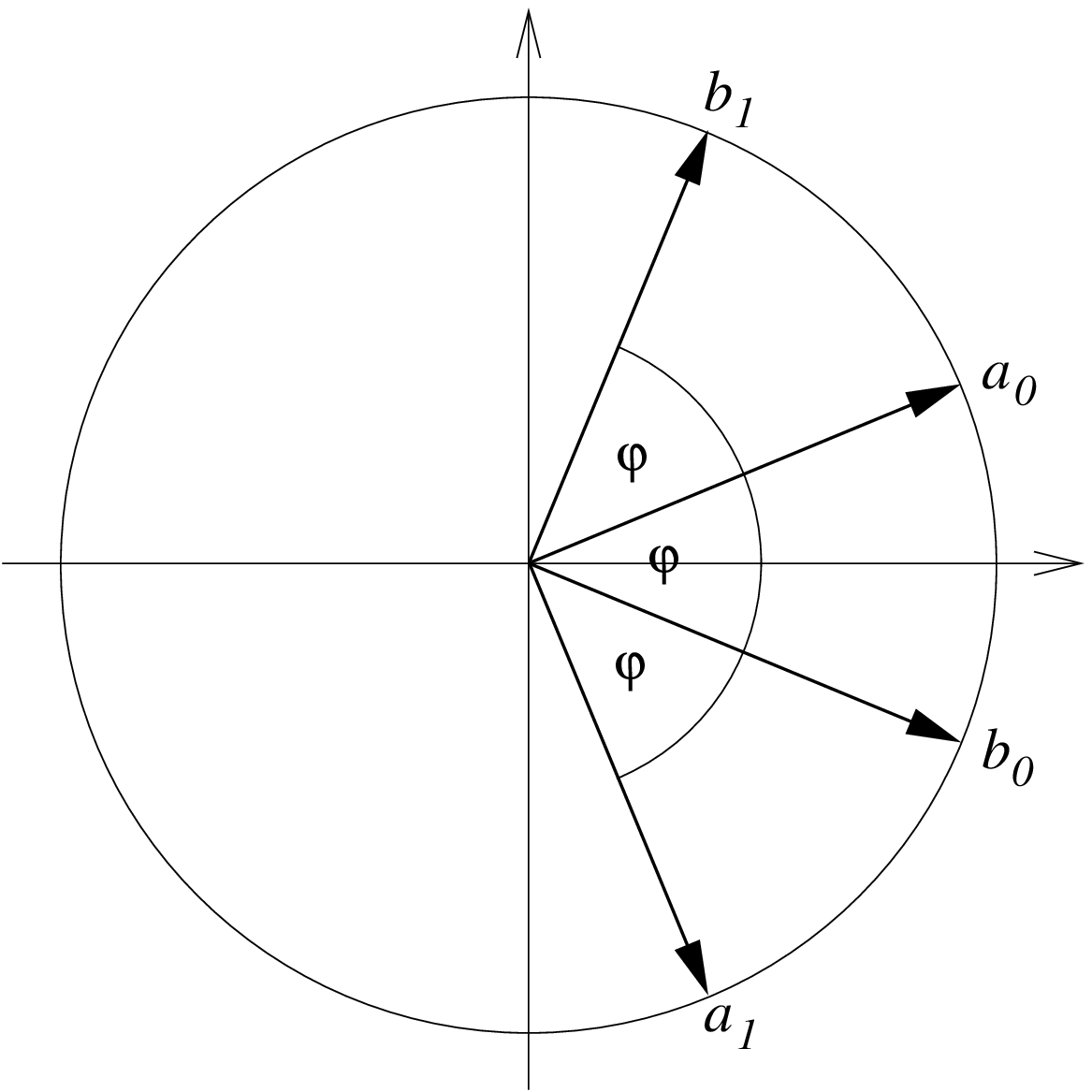}}\qquad\quad
\subfigure[$\frac{2}{3}<\alpha\leq1,\,\varphi=0$]{\includegraphics[width=0.35\textwidth]{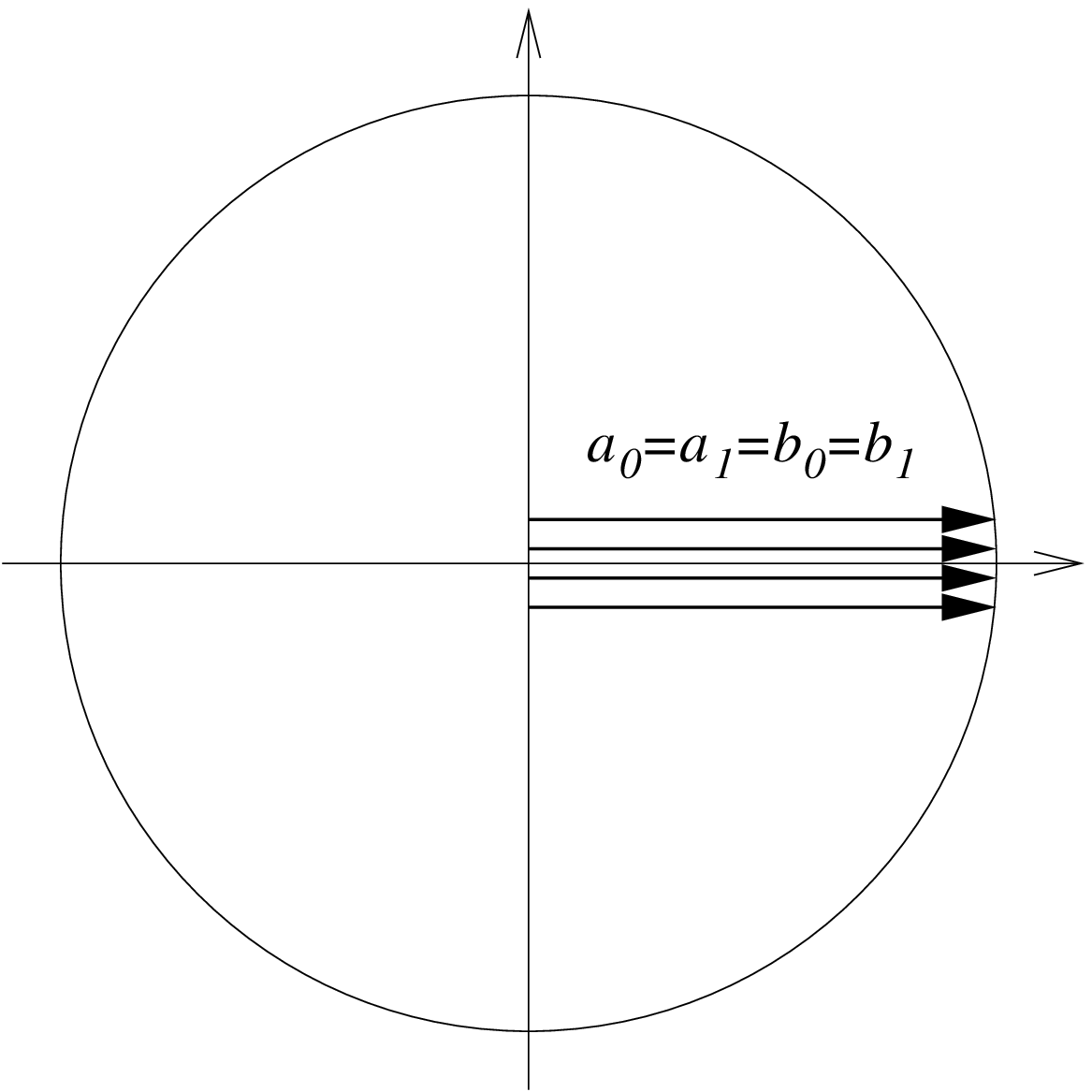}}
\caption{Choice of observables for the maximal violation $I_M(\alpha)$ of the CHSH inequality.} 
\end{figure}

We generalize the previous reasoning about the violation of the CHSH inequality, which we want to maximize for every $0\leq\alpha\leq1$. We follow an experimental procedure like the one depicted above, taking into account that, in principle, the input orthonormal states $\ket{x}_\rmA$, $x=0,1$, and $\ket{y}_\rmB$, $y=0,1$, on Alice's and Bob's sides are arbitrary and independent of the POVM entering in the definition of $\Lambda_{\rm NL}$. This means that we have
\beq
p=|\braket{11}{\psi^{ij}_{\rm in}}|^2=|\braket{11}{i_\rmA j_\rmB}|^2=p^i_\rmA p^j_\rmB,
\eeq
with $p^i_\rmA=|\braket{1}{i}_\rmA|^2$ and $p^j_\rmB=|\braket{1}{j}_\rmA|^2$.
Defining
\beq
c_{ij}=(1-\alpha p^i_\rmA p^j_\rmB)\vec{a}_i\cdot\vec{b}_j+\alpha p^i_\rmA, p^j_\rmB\vec{a}_i\cdot\vec{b}^r_j
\eeq
with~\footnote{It is sufficient to consider vectors in the plane $xz$.}
\beq
\vec{a}_i=\big((a_i)_x,0,(a_i)_z\big),\qquad
\vec{b}_j=\big((b_j)_x,0,(b_j)_z\big),\qquad
\vec{b}^r_j=\big(-(b_j)_x,0,(b_j)_z\big),
\eeq
we may write the quantity we want to find as 
\beq
I_{\rm M}(\alpha)=\max_{p^i_\rmA,p^j_\rmB,\vec{a}_i,\vec{b}_j}\{c_{00}+c_{01}+c_{10}-c_{11}\}.
\eeq
For each fixed $\alpha$, $I_{\rm M}(\alpha)$ is attained by
\beq
p^0_\rmA=p^0_\rmB=0,\quad p^1_\rmA=p^1_\rmB=1
\eeq
and 
\begin{align}
\vec{a}_0&=\big(\cos\frac{\varphi}{2},0,\sin\frac{\varphi}{2}\big),&
\vec{b}_0&=\big(\cos\frac{\varphi}{2},0,-\sin\frac{\varphi}{2}\big),\\
\vec{a}_1&=\big(\cos\frac{3}{2}\varphi,0,-\sin\frac{3}{2}\varphi\big),&
\vec{b}_1&=\big(\cos\frac{3}{2}\varphi,0,\sin\frac{3}{2}\varphi\big),
\end{align}
choosing $\varphi$ to maximize
\beq
c_{00}+c_{01}+c_{10}-c_{11}=3\cos\varphi-\cos3\varphi+\alpha(1+\cos3\varphi).
\eeq
The optimal angle is (see Figure \ref{fig:optob})
\beq
\varphi(\alpha)=\begin{cases}
\arcsin\sqrt{\frac{2-3\alpha}{4(1-\alpha)}}& 0\leq\alpha\leq\frac{2}{3}\\
0& \frac{2}{3}<\alpha\leq1
\end{cases}
\eeq
so that
\beq
I_{\rm M}(\alpha)=\begin{cases}
\sqrt{\frac{(2-\alpha)^3}{1-\alpha}}+\alpha& 0\leq\alpha\leq\frac{2}{3}\\
2(1+\alpha)& \frac{2}{3}<\alpha\leq1.
\end{cases}
\eeq

We consider also the family of maps
\beq
\Lambda'_{\alpha}[\rho_{\rm in}]=(1-\alpha p)\frac{\ket{00}\bra{00}+\ket{11}\bra{11}}{2}+\alpha p\frac{\ket{01}\bra{01}+\ket{10}\bra{10}}{2};
\eeq
for $\alpha=1$ we recover the map \eqref{eq:nl_qc_b_2_prime}, while for $\alpha=0$ we get a constant map with separable output.
Looking for the maximal violation of CHSH inequality with our experimental procedure, we have in this case 
\beq
I'_{\rm M}(\alpha)=\max_{p^i_\rmA,p^j_\rmB,\vec{a}_i,\vec{b}_j}
\{d_{00}+d_{01}+d_{10}-d_{11}\}
\eeq
with
\beq
d_{ij}=(1-2\alpha p^i_\rmA p^j_\rmB)(a_i)_z (b_j)_z.
\eeq
We find $p^0_\rmA=p^0_\rmB=0,p^1_\rmA=p^1_\rmB=1$ and $\vec{a}_i=\vec{b}_j=(0,0,1)$, so that
\beq
I'_{\rm M}(\alpha)=2(1+\alpha).
\eeq

\begin{figure}
\centering
\subfigure[]{\label{fig:violbell}\includegraphics[width=0.4\textwidth]{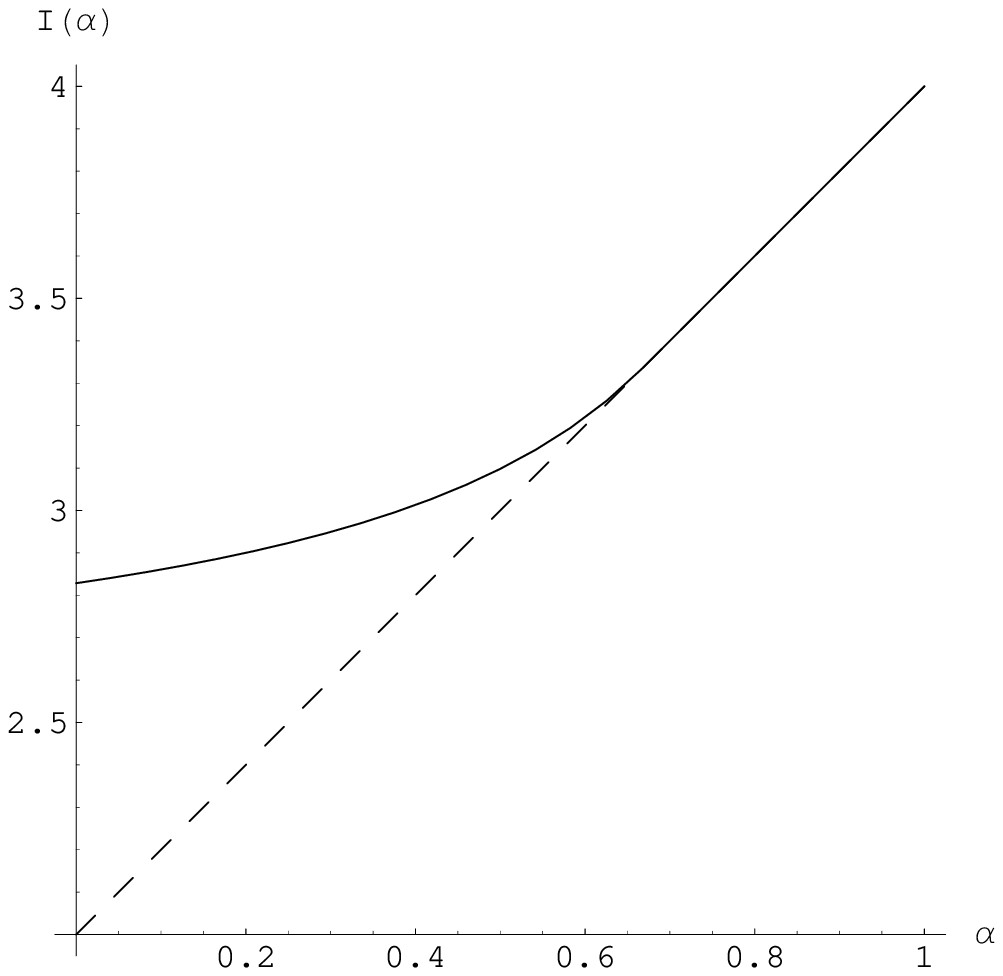}}\qquad
\subfigure[]{\label{fig:violbell2}\includegraphics[width=0.4\textwidth]{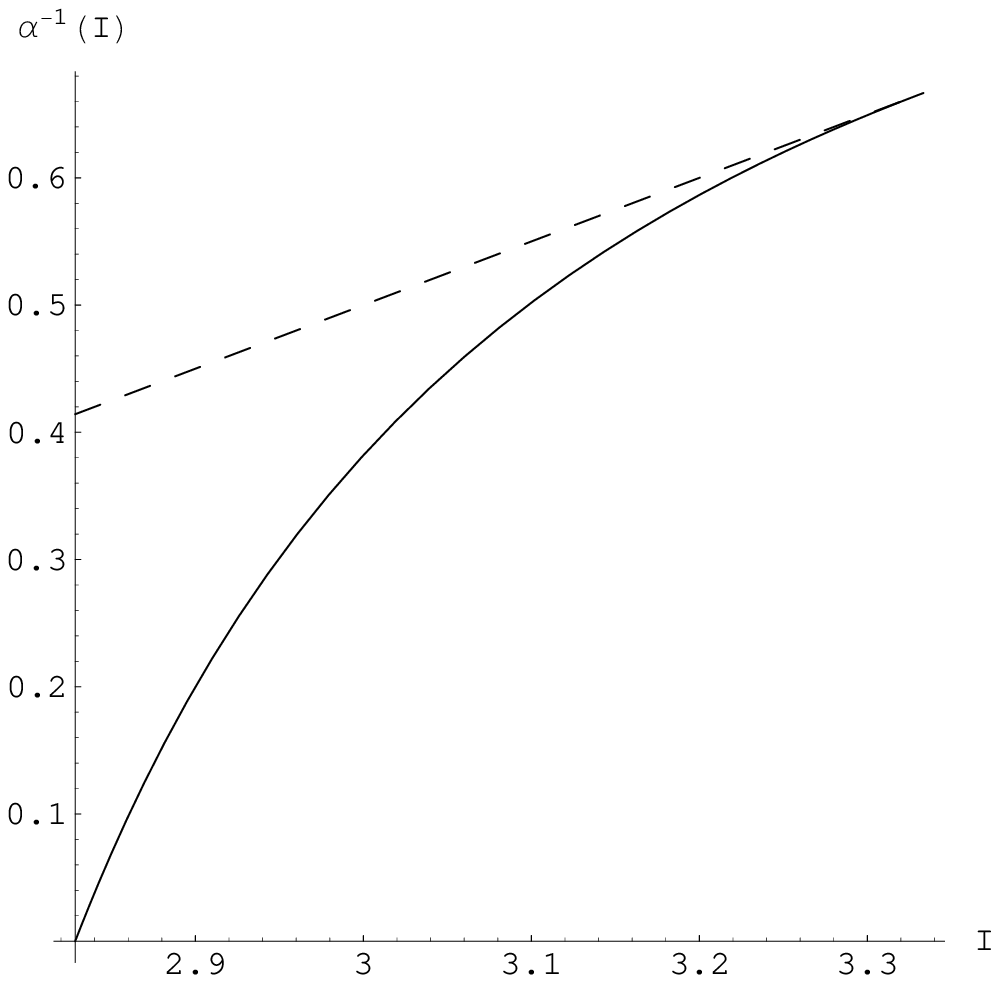}}
\caption{Maximal violation of the CHSH inequality: (a) the coherent quantum box $\Lambda_\alpha$ (continuous line) allows a greater maximal violation of the CHSH inequality with respect to $\Lambda'_\alpha$ (dashed line); (b) less communication appears to be required by the coherent box  to obtain the same amount of maximal violation.} 
\end{figure}

In Figure \ref{fig:violbell} it is shown how, for fixed $\alpha$, the map $\Lambda_\alpha$ allows in general a violation of the CHSH inequality greater than the one obtained through the map $\Lambda'_\alpha$. The maximal violation becomes equal for the two maps in the range $2/3\leq \alpha \leq1$. Conversely, as can be more appreciated in Figure \ref{fig:violbell2}, the map $\Lambda_\alpha$ allows the same amount of violation of the CHSH inequality  than the map $\Lambda'_\alpha$  with lower $\alpha$, i.e., considering the scheme depicted above to realize $\Lambda_\alpha$, with a lower amount of communication. Entanglement can therefore be interpreted as a resource to reduce the communication complexity of violating the CHSH inequality. It is interesting it seems to be of no help for $2/3\leq \alpha \leq1$; in such range of values, the non-locality, as measured by $I_M(\alpha)$, appears to be essentially due to the communication involved in the realization of the operation.

\subsection{Trade-off between nonlocality and entangling power for non-signalling boxes}

\begin{figure}
\centering
\includegraphics[width=0.6\textwidth]{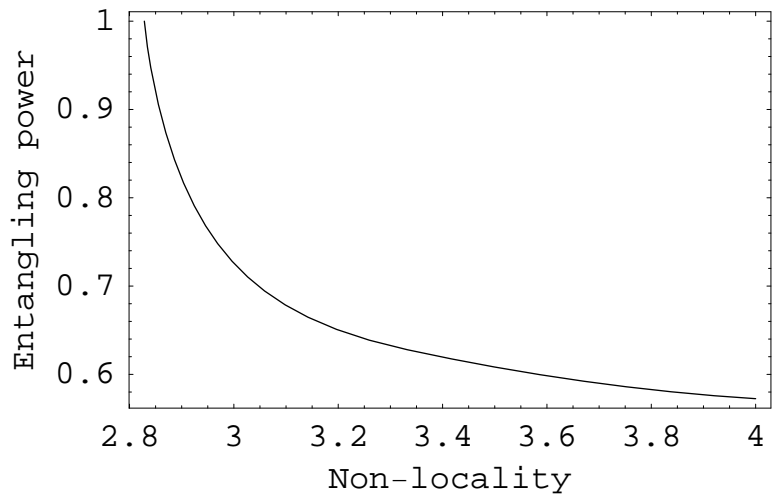}
\caption{Parametric plot of the entangling power versus nonlocality for the family of maps $\Lambda_\alpha$} 
\label{fig:parplot}
\end{figure}

One may define the \emph{entangling power} of a map $\Lambda$ with respect to a measure of entanglement $E$ as
\cite{Zanardi-entpower}
\beq
E_{\rm pow}(\Lambda)\equiv\big\langle E\big(\Lambda[\pro{\psi_1}\otimes\pro{\psi_2}]\big) \big\rangle_{\psi_1,\psi_2}
\eeq
where the average is taken with respect to the initial separable pure state $\pro{\psi_1}\otimes\pro{\psi_2}$, i.e., in general,
\beq
\langle A(\psi_1,\psi_2) \rangle_{\psi_1,\psi_2}=\int d\psi_1 d\psi_2 p(\psi_1,\psi_2)A(\psi_1,\psi_2),
\eeq
with $p(\psi_1,\psi_2)$ a probability distribution over separable pure states.
Since we are considering a map acting on qubits, we adopt the Bloch sphere parametrization
\beq
\label{eq:bloch}
\ket{\psi_i}=\cos\frac{\theta_i}{2}\ket{0}+e^{i\phi_i}\sin\frac{\theta_i}{2}\ket{1},
\eeq
with $0\leq\theta_i<\pi,\,0\leq\phi_i< 2\pi$, and the corresponding uniform probability distribution
\beq
d\psi_1 d\psi_2 p(\psi_1,\psi_2)=\frac{1}{(4\pi)^2}d\phi_1 d\phi_2 d\theta_1 d\theta_2 \sin\theta_1 \sin\theta_2.
\eeq
As entanglement measure we will adopt the relative entropy of entanglement~\cite{VPRK1997,PlenioVedral1998}, $E=E_{\rm R}$; 
it takes an easy expression in the case of a Bell diagonal state $\rho_{\rm Bell}$ of two qubits~\cite{VPRK1997,PlenioVedral1998}: 
\beq
E_{\rm R}(\rho_{\rm Bell})=
\begin{cases}
1-H(\lambda)	& \lambda>1/2\\
0			& \lambda\leq1/2,
\end{cases}
\eeq
where $\lambda$ is the highest eigenvalue of $\rho_{\rm Bell}$ and $H(x)$ is the binary Shannon entropy
\beq
H(x)=-x\log x-(1-x)\log(1-x).
\eeq
In our case the output of $\Lambda_\alpha$ is always a rank-two Bell diagonal state, so that
\beq
E_{\rm R}(\Lambda_\alpha[\rho])=1-H(p)
\eeq
with $p=\bra{11}\rho\ket{11}$.
Given the parametrization \eqref{eq:bloch}, we have
\beq
p=p(\theta_1,\theta_2)=\sin^2\frac{\theta_1}{2}\sin^2\frac{\theta_2}{2},
\eeq
so that
\beq
\begin{split}
E_{\rm pow}(\Lambda_\alpha)&=
1-\frac{1}{(4\pi)^2}\int_0^{2\pi}\!\!\!\! d\phi_1 \int_0^{2\pi}\!\!\!\! d\phi_2  \int_0^\pi \!\!\!\!d\theta_1 \int_0^\pi\!\!\!\! d\theta_2 \sin\theta_1 \sin\theta_2\;  H\Big(\alpha\sin^2\frac{\theta_1}{2}\sin^2\frac{\theta_2}{2}\Big)\\
&=1+\frac{1}{\alpha}\int_0^{\sqrt{\alpha}}\!\!\!\!du\int_0^{\sqrt{\alpha}}\!\!\!\!dv\Big(
uv\log uv+(1-uv)\log(1-uv)\Big).
\end{split}
\eeq
In Figure \ref{fig:parplot} the entangling power $E_{\rm pow}(\Lambda_\alpha)$ is plotted against the degree of violation of the CHSH inequality $I_{\rm M}(\alpha)$. The two quantities are inversely correlated: the more the degree of non-locality, the lesser the average amount of entanglement created.

Thus we have obtained an example of trade-off capability of creating
nonlocal correlations
and creating entanglement. The box that violates Bell's inequalities
in the strongest way, is not so good in producing entanglement
out of product states on average. This can be explained by the
fact that if an operation is to be strongly nonlocal,
and yet, is not to allow signalling, it must produce some
masking noise.

Note that we can find an analogue of such behaviour also
for classical boxes. Obviously, we see that in PR box,
the outputs, apart from satisfying equation $a\oplus b=xy$
are otherwise completely random, to ensure the property of non-signalling.
Now, let us discuss how such randomness (i.e. noise) implies a
trade-off analogous to that between nonlocality and entangling power.
Namely, instead of entangling power we can think of the ``correlating power''
defined as follows. We first choose some measure $C$ of correlations
for classical states. Then the quantity is defined in a similar way
as entangling power: it is the average output $C$
produced out of initial product input states by the box. Thus $C$ plays here
a role
analogous to that of an entanglement measure in the definition of the entangling power.
It is not hard to agree that for classical states,
the uniformly, perfectly correlated state must have maximal $C$,
if only we assume that $C$ does not increase under local operations,
which is quite reasonable for correlation measure \cite{HendersonVedral}.
Thus, if we take a box that replaces any state with maximally
correlated one, it will have maximal possible correlating power.
On the other hand, PR box cannot all the time produce maximally correlated
states.
Thus again, the price for being nonlocal is smaller capability of creating
correlations.

It is interesting though, that in quantum case there is some effect
that does not have analogue in classical case. Namely, we can consider ``power of producing classical correlations''. There can be many measures
of classical correlations contents of quantum state
(cf.
\cite{HendersonVedral,IBMHor2002,DevetakW03-common,compl,SynakH04-deltacl,igor-deficit}).
Nevertheless it is plausible to assume, that
a quantum state has maximal possible classical correlations
when there exists local measurements that produce classical, maximally
correlated
state. This is exactly the situation with quantum nonlocal box,
since it always produces a mixture of two singlets.
Thus the power of producing classical correlations
is the maximal possible one and it does not go down with increasing nonlocality of the
box.

\section{Boxes, bound entanglement and Carnot reversibility}
\label{sec:bound}

In this section we will discuss irreversibility exhibiting by boxes in analogy to bound entanglement. 
The latter was connected with thermodynamics \cite{thermo-ent2002}, and a sort of Carnot reversibility was found to be impossible in general. 
Below we analyze this issue for (mainly classical) boxes, basing on the result of van Dam \cite{vanDam-compl}  on communication 
complexity and nonlocal boxes.

\subsection{Carnot reversibility and entanglement}
Bipartite bound entangled states are states from which one cannot distill singlets 
by local operations and classical communication. However they are entangled, so that to create them, one needs singlets \cite{bound,Vidal-irrev2001}. In general, for many states we have that the amount of singlets one can draw is smaller that the amount needed
to create them. In other words, entanglement cost $E_C$ is usually greater than distillable entanglement  $E_D$, and for 
bound entangled states $E_C>0$ and $E_D=0$ \cite{note-reg}. The difference between $E_C$ and $E_D$ 
can be called content of bound entanglement in the state, $E_B=E_C-E_D$. For bound entangled states the whole entanglement 
is bound $E_C=E_B$. Thus there are two types of entanglement: ``free'' entanglement, that can be used to perform quantum communication,
and bound entanglement. This was compared with thermodynamics, where we have two types of energy: free energy, capable to perform 
work, and bound energy. In a sense bound entanglement is like a \emph{single} heat bath, which though contains energy, 
cannot be used for drawing work cyclically. Irreversibility between formation and distillation is analogous to irreversibility 
exhibited in the process of dissipation, where e.g. mechanical energy is changed into heat. 

We know that in thermodynamics, apart  from irreversibility, we have also reversibility in  Carnot cycle. Carnot cycle
shows that it is possible to dilute work into heat and get it back. In \cite{nlocc,uniqueinfo} we have shown that similar reversibility holds for purity and noise.
That is changing pure state into mixed one is irreversible, but one can {\it regain} reversibility, if one creates a given state 
with initial maximally mixed states and pure states in proper proportions. Thus purity can be reversibly diluted into noise, like work into heat. 
In \cite{thermo-ent2002} we have analyzed whether this is possible 
in entanglement theory, i.e. whether singlets can be diluted into bound entanglement reversibly.
It would mean that it is possible to 
reversibly obtain any state from singlets (analogue of work) and bound entangled states (analogue of heat). 
The amount of needed singlets to create $\rho$ would be equal to $E_D(\rho)$. Apart from singlets, one would use some bound entangled state 
$\sigma_{b}$ with $E_C(\sigma_b)=E_B(\rho)$. Thus the singlets would contribute to free entanglement only, while bound entangled state $\sigma$ 
would contribute to bound entanglement. 
We have found that up to unproven, but plausible assumptions, it is not the  case in general (cf. \cite{APE}).

Before we pass to discuss boxes, let us mention that we have a hierarchy of sets in entanglement theory. 
There is the set of separable states, which can be created solely by local operations and classical communication (LOCC),
without singlets (equivalently without quantum communication). These states thus do not represent a resource in a LOCC framework, we can have them for free. 
Then we have the set of bound entangled states, which cannot be created for free,  but can not be distilled  (through LOCC) to give singlets. Finally there is ``the most valuable''
set of states, distillable states.

\subsection{Quantum boxes and Carnot reversibility }

Let us discuss the issue of boxes in this context. First of all let us say what is the resource, and 
what is for free. The resource is in this case communication, whether classical or quantum. The states, whether product, 
separable or entangled are for free. Also local operations are public resource. Therefore we may consider the communication resource to be only classical, the quantum one being obtainable via teleportation.
  
The analogue of worthless separable states are localizable boxes. They can be produced 
for free, i.e. without communication (though perhaps using entangled states as ancillae). 
Then there is set of causal boxes, analogue of nondistillable states. Such boxes cannot be used to communicate. 
The boxes that are causal, but are not localizable, are analogous to bound entangled states: 
one needs communication to implement them, however they do not ``return'' the communication, it is lost.
Finally, for a generic box, one expects that the amount of communication needed to implement the box, $C_C$,
will usually exceed the amount of communication that can be obtained by use of the box, $C_D$. Thus we can speak about 
bound, $C_B=C_C-C_D$, and free communication. In the following we will consider classical boxes, and in the end of the subsection we will go back 
to quantum boxes.

Let us discuss the possibility of Carnot reversibility. For boxes it would mean that any box could be built 
out of causal boxes and communication, in such a way that the amount of used communication 
is equal to the communication that we can get back from the box. For example consider inner product 
\beq
f(x,y)=\sum_i x_i y_i 
\eeq 
where $x$ and $y$ are $n$-bit strings, respectively in Alice's and Bob's hands, and addition is modulo 2. The function has one bit output 
(to make it a two-bit output map, we could for example return $f(x,y)$ to Bob and a fixed bit to Alice).
It is known that to perform such box $n$ bits of communication are needed \cite{vanDam-n}, i.e. there is no better way 
than the following:  Alice sends all bits to Bob, 
and Bob computes the value $f(x,y)$. The amount of communication that can be obtained 
from this box cannot be of course more than one, being the output (on Bob's side) one bit. Thus we have large irreversibility, the ``bound  communication'' 
amounting to $n-1$ bits. 

Now, we ask if we  can gain reversibility when we implement this box out of causal boxes and communication.
We can use a result by van Dam \cite{vanDam-compl}, who shows that by use of $n$ PR-boxes (each has at most 1 bit of bound communication)
inner product can be computed at additional cost of only 1 bit of communication. 
Thus in this case there is no loss of free communication. It is interesting then to ask the following question: {\it Given that causal boxes are a public resource,
what boxes can be implemented by use of the same amount of communication that can then be recovered from them?}

Actually, it turns out, that any function with  $(n+n)$-bit input and 1 bit output, 
can be realized by $n$ PR boxes plus a single bit of communication \cite{vanDam-compl}. If such a function 
can be used for one bit of communication (like inner product), then we do not lose free communication. This happens as soon as the function to be computed is not \emph{independent} of the input of the party which we consider the sender of the message (in our case Alice). Indeed, if it is independent then the \emph{total} communication needed to compute the function is zero (neither causal boxes nor free communication are involved in the process); if it is not, then for sure there is a fixed input configuration of the receiver, Bob, such that the Alice can send a 1-bit-message changing her input. 

Note that our discussion serves also as interpretation of van Dam result. The fact that causal boxes can be so helpful, 
can be interpreted as follows: the functions that have 1 bit output, have a lot of bound communication. Thus they are themselves 
almost like causal boxes. Thus it is less surprising, that 
one can build such function out of many causal boxes and small communication. What remain surprising, is the 
very existence of causal boxes characterized by $C_C>0$ and $C_D=0$.

For quantum boxes, the question of Carnot reversibility appears to be more involved. We can ask if there is analogous result to that of van Dam: can quantum communication complexity 
be reduced to a single qubit for boxes with 2n-qubit input and 1-qubit output, if causal boxes are public resource?

\section{Conclusions}
We have provided new results on non-signalling boxes, and alternative proofs
for existing results. We have shown that causal boxes are rare in the set of all maps. Then we have 
investigated connection of causal maps that are not localizable with the set of entanglement 
breaking maps. An open question is whether there exist maps that are causal,
but do not belong to convex hull of causal entanglement breaking maps and localizable maps. We have related the ``construction''/classification of causal maps to the study of classes of states with fixed reductions. Subsequently, 
we have found a trade-off between nonlocality and entangling power for a family of non-signalling maps. 
It would be interesting to prove such a trade-off for the whole family of causal maps: indeed, 
it must be that the very condition of causality imposes such  a trade-off.
Finally, we have related the issue of irreversibility for causal boxes (they need communication,
but do not return it) to a similar one in entanglement theory and 
the result of \cite{vanDam-compl}, that with use of non-signalling boxes  distant parties 
can compute any 1 bit valued function 
by use of one bit of communication, with thermodynamical reversibility,
that was sought to present entanglement theory \cite{thermo-ent2002}. 

As far as new proofs are concerned, we have provided a compact proof of 
theorem of \cite{beckman} characterizing semicausal maps, of the fact that nonlocal unitaries 
are signalling, and that semicausalilty and semilocalizability are equivalent. 

There are many interesting questions that one can pose in the context of non-signalling maps. 
One of them is the following: can causal boxes play a role in design of fault tolerant 
schemes? The reason why they could be useful, is that because of causality,
they do not propagate errors. So one could try to design circuits in such a way, 
that if only possible, a causal box is put instead of usual gate. 
Of course, the fact that the causal boxes to not propagate errors, 
means that they also do not propagate information, hence it is not clear, 
if they can be useful elements of circuits.

Another important question concerns complexity of distributed quantum computing. 
Can we have quantum analogue of van Dam result, i.e. can any map with one qubit output 
be implemented by causal boxes and only one qubit of communication?

Still the basic problem of characterizing the set of causal boxes in terms of extremal points 
remains open. We hope we have provided here an interesting direction by relating it to 
the issue of entanglement breaking maps. 

We would like to thank Nicolas Gisin, Caterina Mora, Jonathan Oppenheim and
Sandu Popescu
for stimulating discussions. The work has been supported by the Polish
Ministry of Scientific
Research and Information Technology under the (solicited) grant No.
PBZ-MIN-008/P03/2003 and by EC grants RESQ, Contract No. IST-2001-37559
and QUPRODIS, Contract No. IST-2001-38877.

\appendix*

\section{Convex analysis}
\label{app:convex}
 
 Given $V$, a liner space over $\mathbb{R}$, a line segment with endpoints $\vec{x}$ and $\vec{y}$ in $V$ is the set
\beq
[\vec{x},\vec{y}]=\{\lambda\vec{x}+(1-\lambda)\vec{y}|0\leq\lambda\leq1\};
\eeq 
the interior of such a segment, when  $\vec{x}\neq\vec{y}$, is the set
\beq
]\vec{x},\vec{y}[=\{\lambda\vec{x}+(1-\lambda)\vec{y}|0<\lambda<1\}.
\eeq 
A line through two points $\vec{x},\vec{y}\in V$ is defined as
\beq
\langle\vec{x},\vec{y}\rangle=\{\lambda\vec{x}+(1-\lambda)\vec{y}|\lambda\in\mathbb{R}\}.
\eeq
The algebraic interior $W^i$ of a subset $W\subseteq V$ is given by all $\vec{w}\in W$ such that for every line $l$ through $w$, the intersection $l\cap W$ contains a segment having $w$ in its interior.


\begin{thebibliography}{35}
\bibitem[{not()}]{note-reg}
\bibinfo{note}{Note however, that bound entangled states are defined as those
  that are not distillable, but are entangled. It is not known, if all such
  states in the asymptotic limit need singlets to be created, though it is
  highly plausible. Thus so far we do not know if for all bound entangled
  states $E_C>0$. See \cite{Vidal-cost2002} in this context.}
\end{thebibliography}

\end{document}